# Bar-driven Fueling to a Galactic Central Region in a Massive Gas Disk

## Keiichi Wada,[1] Asao Habe[2]

[1] *Center for Information Processing Education, Hokkaido University, Kita 10 Nishi 5, Sapporo 060, Japan*
*e-mail: wada@hipecs.hokudai.ac.jp*
[2] *Department of Physics, Hokkaido University, Kita 10 Nishi 8, Sapporo 060, Japan*
*e-mail: habe@phys.hokudai.ac.jp*

**ABSTRACT**

We have found an effective fueling process to a central region of galaxies with weak bar-like distortion by two dimensional hydrodynamical simulations. Gravitational instability of an elongated gas ring at the inner Lindblad resonance (ILR), which has been reported as an effective fueling mechanism, are not needed for this fueling process. A massive gaseous disk in a central region of galaxies sensitively responds to the weakly distorted potential, and a large amount of gas can be fed into within 1/20 of a core radius of the potential in several $10^7$ yr. The ILRs, the dissipative nature of the gas, and self-gravity of the gas are essential for triggering this effective fueling. The accumulation process has not been ever known: the gas accumulates to form a dense 'linear' structure inclined at about 45 degree with respect to the bar potential in a *leading sense*. We also found that a counter rotating gaseous core can be formed as a result of the fueling. The sense of the rotation of the core depends on a fraction of the gas mass to the background mass. Physical mechanism of the fueling process can be understood using a linear theory of gaseous orbits in a weak barred potential.

## 1 INTRODUCTION

It is generally accepted that nuclear activities in galaxies, such as starbursts, result from feeding of the gas into a nuclear region (e.g., see a review by Phyinney 1994). Many studies in the last two decades have revealed the fueling process from 10 kpc to hundred pc from the center. A stellar bar, which is formed by gravitational instability of a stellar disk or triggered by close-encounters of galaxies (e.g., Barnes and Hernquist 1992), is the most probable cause of the fueling. Due to angular momentum exchange between the gas and the bar, a large amount of gas can be accumulated near to the inner Lindblad resonances (ILR), which is usually located at $\sim 1$ kpc from the center. Numerical simulations, however, have cleared that the bar cannot force the gas to fall beyond the ILR. To break the 'ILR barrier', self-gravity of the gas is essential (Fukunaga & Tosa 1991; Wada & Habe 1992 (hereafter Paper I); Elmegreen 1994). Gravitational instability followed by collapse of the extremely elongated ILR ring results in a rapid mass accretion, $\sim 10^9 M_\odot$ in $5 \times 10^8$ yr to several hundred pc. However, fueling from several hundred pc to the central 1 pc needs a different mechanism: secondary a gaseous or a stellar bar (Shlosman, Frank, Begelman 1989; Friedli, Martinet 1993) or effective viscosity caused by an active star formation and supernovae explosions (Yamada 1994) were proposed.

In the present paper, we have investigated dynamical evolution of a massive gaseous disk in a central region of a galaxy with weak distortion. Massive gas disks should have been common in the galaxy-formation era as well as in nearby starburst galaxies (Turner 1994). Recent near infrared observations have often revealed weak bar-like structure in the central region of normal spiral galaxies (e.g., Block & Wainscoat 1992). By two dimensional hydrodynamical simulations taking into account the self-gravity of the gas, we have found a new fueling mechanism from kpc to several tens pc, which is induced by a weak bar and the self-gravity of the gas. Preliminarily results have been reported by Wada and Habe (1994). A large fraction of a massive gaseous disk inside the inner ILR is fed into the central 100 pc region in a time scale of several $10^7$ yr. We have also found that a dense nuclear core formed by this process shows a counter rotation relative to the galactic rotation under a certain condition. Our result suggests that the merger of galaxies is not the only way to form the observed counter-rotating core in galaxies.

In the next section, we describe numerical models, ini-tial conditions, and a numerical method. Numerical results are shown in section 3. In section 4, we discuss physical interpretations of the rapid fueling and the counter-rotating core.

## 2    NUMERICAL MODELS AND METHOD

### 2.1    Model galaxy

The numerical method and models are based on Paper I, but with higher numerical resolution. In this paper, we concentrate evolution of a gas disk inner a few kpc region of a galaxy. We assume that the model galaxy is composed of a stellar disk and a bulge component with small distortion. The stellar bar component is treated as an external fixed potential, i.e., time independent in its shape. We do not consider a dark halo component, since we are interested in gas motion in the inner region of the galaxy. Axi-symmetric potential $\Phi_0$ is assumed to be 'Toomre disk' (Toomre 1963) as,

$$\Phi_0(R) = \frac{c^2}{a}\frac{1}{(R^2 + a^2)^{1/2}} \ , \tag{1}$$

where $a$ is a core radius and $c$ is given as $c = v_{max}(27/4)^{1/4}a$, and $v_{max}$ is a maximum rotational velocity in this potential. We chose $v_{max} = 250$ km s$^{-1}$, and $a = 2.0$ kpc.

We assume the barred potential as,

$$\Phi_1(R,\theta) = \varepsilon(R)\Phi_0 \cos 2\theta \ , \tag{2}$$

where $\varepsilon(R)$ is given as,

$$\varepsilon(R) = \varepsilon_0 \frac{aR^2}{(R^2 + a^2)^{3/2}} \ , \tag{3}$$

and $\varepsilon_0$ is a parameter which represents strength of the bar potential to the disk component (Sanders 1977). We have explored models with $\varepsilon_0 = 0.0, 0.05, 0.1$ and $0.2$. Since a maximum ratio of the non-axisymmetric part to the axisymmetric one is $\max(|\Phi_1(R)|/|\Phi_0(R)|) = 0.385\ \varepsilon_0$, our model bars are very weak.

There are three sorts of resonances for orbits in a weak bar potential, $\Omega_b = \Omega_0(R)$ : corotation resonance, $\Omega_b = \Omega_0(R) \pm \kappa_0(R)/2$ : inner$(-)$ and outer$(+)$ Lindblad resonance, where $\Omega_b$ is the bar pattern speed, $\Omega_0(R)$ is a circular frequency given as $\Omega_0^2 = -R^{-1}d\Phi_0/dR$ , and $\kappa_0(R)$ is the epicyclic frequency given as $\kappa_0(R)^2 = Rd\Omega_0^2/dR + 4\Omega_0^2$.

Taking into account the gravitational potential of the uniform gas disk (see the next subsection), the circular frequency is

$$\Omega_0^2(R) = \frac{1}{R}\left[\alpha\frac{R}{(R^2+a^2)^{1.5}} + \beta\right] , \tag{4}$$

where $\alpha \equiv c^2/a$, $\beta \equiv G f_g M_{dyn}/R_{gas}^2$, where $f_g$ is a ratio of gas mass to the dynamical mass, $M_{dyn}$, derived from equation (1). The epicyclic frequency $\kappa_0$ is

$$\kappa_0^2 = \frac{3}{R}\left[\alpha\frac{R}{(R^2+a^2)^{1.5}} + \beta\right] + \alpha\frac{a^2-2R^2}{(R^2+a^2)^{2.5}}. \tag{5}$$

Fig. 1

We plot $\Omega_0 - \kappa_0/2$ for $f_g = 0.17$ and $0.10$ against the radius in Fig. 1. If $0.005 < \Omega_b < 0.026$ km s$^{-1}$ pc$^{-1}$, there are two ILRs for $f_g = 0.10$ inside the initial gas disk (the radius is 2.0), whereas there is one ILR at the vicinity of the center if $\Omega_b > 0.026$, and no ILRs for $\Omega_b < 0.005$. We mainly explored models with $\Omega_b = 0.02$, since we are interested in behavior of the gas near the center and the role of the ILR, and the dust lanes in bars implies that the bar has at least one ILR (Sanders & Tubbs 1980; Athanassoula 1992). In paper I, we found that gas disks in a weak bar with high pattern speed, for which there are no ILRs, does not show any accumulation process near the center. The ILRs are essential for the gaseous dynamics in galaxies (we discuss this subject in section 4.1).

We take the following units in this paper; $[T] = 10^7$ yr, $[L] =$ kpc, $[M] = M_\odot$, if we do not explicitly state the unit.

### 2.2    Gas disk models

In Smoothed Particle Hydrodynamics (SPH) code, a gas disk is represented by a large number of 'quasi-particles' with a certain spatial extent. Total number of the particles in our models is $10^4$. The SPH particles are randomly distributed within $R_{gas} = 2.0$ in order to represent uniform density disk at $t = 0$. The initial rotational velocity is given in order to balance centrifugal force caused by the axisymmetric gravitational potential, since the distorted potential is very weak. The rotational periods at $R = 0.2, 1.0$, and $2.0$ for $f_g = 0.15$ are $2.5, 3.4$, and $4.9$, respectively. We assume that gas is isothermal (the sound velocity is $0.1$).

Besides the rotational velocity of the potential and the bar-strength ($\varepsilon_0$), a fraction of the gas mass to the dynamical mass ($f_g$) is an important parameter. We assume $f_g = 0.03 - 0.30$. Total dynamical mass inside the gas disk, $M_{dyn}(R_{gas})$ is $2.3 \times 10^{10} M_\odot$. Models are summarized in Table 1.

### 2.3    Numerical method

The equations of motion of $i$-th SPH particles are,

$$\frac{d\boldsymbol{r}_i}{dt} = \boldsymbol{v}_i, \tag{6}$$

and

$$\frac{d\boldsymbol{v}_i}{dt} = -\nabla(\Phi_{axi} + \Phi_{bar}) - \nabla\Phi_{gas} - \frac{1}{\rho_i}\nabla(P_i + q_i) \ , \tag{7}$$

where $\boldsymbol{r}_i$ is the position vector of $i$-th particle from the center of disk, and $\Phi_{axi}$ and $\Phi_{bar}$ are axi-symmetric disk potential and the barred potential defined in the previous sub-section, respectively. $\Phi_{gas}$, $P_i$ and $q_i$ are gravitational potential of gas, gas pressure, and artificial viscosity, respectively. Since we do not consider vertical structure of the gas disk, i.e., we assume a 2-D disk, the forces in equation (7) are calculated in only gas disk plane. The equations of motion (6) and (7) are integrated by a second ordered time centered scheme.

In SPH, the pressure gradient in equation (7) is calculated from following equation,

$$\frac{1}{\rho}\nabla P = \nabla\left(\frac{P}{\rho}\right) + \frac{P}{\rho^2}\nabla\rho \ , \tag{8}$$

where gradient of a physical variable $Q$ at $\boldsymbol{r}$, is calculated as,

$$\nabla(\rho Q) = \sum_{i=1}^{N} m Q_i \, \nabla W(|\boldsymbol{r}_i - \boldsymbol{r}|) \; . \tag{9}$$

The kernel function $W(|\boldsymbol{r}_i - \boldsymbol{r}|)$ is,

$$W(|\boldsymbol{r}_i - \boldsymbol{r}|) \equiv \frac{1}{\pi^{3/2} h^3} \exp\left(-\frac{|\boldsymbol{r}_i - \boldsymbol{r}|^2}{h_i^2}\right), \tag{10}$$

where $m$ is mass of a particle and the particle 'size' $h_i$ is varied depending on the local gas density as $h_i \propto \rho_i^{1/3}$.

The artificial viscosity is given by

$$\frac{1}{\rho_i} \nabla q_i = \sum_j \left(\frac{P_i}{\rho_i^2} + \frac{P_j}{\rho_j^2}\right)(-\alpha\mu_{ij} + \beta\mu_{ij}{}^2)\nabla W(|\boldsymbol{r}_i - \boldsymbol{r}_j|), \tag{11}$$

and

$$\mu_{ij} \equiv \frac{2h\boldsymbol{v}_{ij} \cdot \boldsymbol{r}_{ij}}{c_s(r_{ij}{}^2 + 0.1h_i^2)}, \alpha = 0.5, \beta = 1.0, \tag{12}$$

where $c_s$ is a sound velocity, $\alpha$ and $\beta$ are chosen to represent a shock wave.

Time step, $\Delta t$, is determined by

$$\Delta t = \xi \min\left[\min\left(\frac{h_i}{v_i}\right), \frac{\min(h_i)}{c_s}\right], \tag{13}$$

where we chose a constant $\xi = 1.0$.

The gravitational potential of gas is calculated by solving Poisson's equation,

$$\Delta \Phi_{\text{gas}} = 4\pi G \rho_{\text{gas}}. \tag{14}$$

The gravitational potential $\phi_{i,j}$ at a mesh point $(i,j)$ can be written,

$$\phi_{i,j} = \sum_{i',j'} \mathcal{G}_{i-i',j-j'}\rho_{i',j'}, \tag{15}$$

where $\mathcal{G}$ is the Green's function of point mass gravity. With the convolution theorem, the local potential in equation (14) can be written as

$$\hat{\phi}_{i,j} = \hat{\mathcal{G}}_{i,j}\hat{\rho}_{i,j} \; , \tag{16}$$

where the circumflex denotes the coefficient of Fourier Transform. In order to get the Fourier Transform, we use the FFT (First Fourier Transform) method. If we make an appropriate choice of Green's function, and prepare to use one-quarter of available mesh points for the source distribution, the convolution method enable the potential to be found for an isolated source distribution (Hockney & Eastwood 1981). We use $256 \times 256$ Cartesian grids and the size of grids is 0.05.

## 3 RESULTS

### 3.1 Rapid Fueling

We have found extremely non-axisymmetric gaseous inflow in most models. Figure 2(a) shows typical evolution of the gas disk (model 37) in the co-rotating frame of the bar.

---
Fig. 2(a) (b) (c)
---

A dense linear structure, which inclines by about $30°$ with respect to the major axis of the distorted potential, is formed at the center in the first $T \sim 1.5$. The linear dense region becomes denser with time, and the gas in the dense region fall toward the center. As a result, an extremely dense core is formed at center. Size of the core is typically less than 0.1 and the core contains $\sim 1/3$ of the total gas (Fig. 2(c)).

Velocity field of the gas in the same model (Fig. 2(b) and 2(c)) shows more clearly the peculiar dynamics of the gas. In the early stage, the gas moves on elliptic orbits oriented by $30° \sim 45°$ with respect to the oval potential, and the distortion of the orbits becomes strong with time. Eventually shocks are generated along with the major axis of the elliptical orbits. When the gas on elongated orbits rushes into the shocked region, their orbits drastically change and the gas falls toward the center (Fig. 2(c)).

An elongated ring is formed in the outer region simultaneous with the gas infall. Density of the ring is not uniform; two dense region are located near the end of the ring. This elongated ring is the ILR ring reported in previous numerical simulations; The two dense region in the ring eventually grow to clumps and the ring becomes gravitational unstable (Fukunaga & Tosa 1991; Paper I). However, the fueling process to the galactic center as seen Fig. 2 has not been reported. This is because less massive disks ($f_g = 0.007$-$0.014$ in the notation of this paper) were investigated in the previous simulations. One should note that the gaseous distribution as a whole elongates in a direction at about $45°$ leading to the major axis of the distorted potential.

---
Fig. 3
---

Evolution of gas mass in the central region depends on $f_g$ and $\varepsilon_0$. Figure 3 is time variation of the gas mass inside $R = 0.2$, and shows that the fueling process depends on the strength of the bar. The onset of the rapid fueling is later in more weakly distorted models. This feature has also been seen in the fueling process by collapse of the ILR rings (Paper I), and implies that the angular momentum transfer caused by the external non-axisymmetric gravitational field is the essential for triggering the fueling process. About $1/3$ of the initially distributed gas is fed into the central region in $T = 1$. This accretion rate correspond to $52 M_\odot$ yr$^{-1}$ for model 37 and 47 ($f_g = 0.15$). On the other hand, very weak bar model 22 ($\varepsilon = 0.05$ and $f_g = 0.15$) does not show the rapid fueling. This is because gas accumulate to form a dense lineal structure like as that in the other models, however, the dense region fragments into several clumps, and does not fall to the center over $T = 10$ (Fig. 4). Model 43 ($\varepsilon = 0.1$ and $f_g = 0.15$) is an intermediate case between the rapid fueling models and model 22: the dense linear region fragments to a few clumps, but they eventually merge into a massive core at the center ($T \sim 8.0$).

---
Fig. 4; Fig. 5(a), (b)
---

As shown by Fig. 2, the accretion flow is far from circular. In this case, an isovelocity contour map strongly depends on the position angle of the line of node (Fig. 5(a) and 5(b)). For example, the strong shock, i.e., the large velocity gradient, does not trace the linear dense region of the gas distribution for the model with the position angle $135°$ (Fig. 5(a)). This result implies that if there is strong non-circular motion of the inter stellar matter in galaxies, one should be careful to interpret observational isovelocity maps of these galaxies. Also one should remember, strong non-circular motion can be induced in even galaxies with very weak distortion of the stellar distribution ($f_g = 0.08 - 0.12$: models 55, 47 and 45). Very weak distortion of the potential ($\varepsilon_0 = 0.05$) which corresponds to the axial ratio of the potential $\sim 0.9$ is enough for causing the strong non-circular motion in cooperation with the self-gravity of the gas. On the other hand, no gas accumulation occurs in a completely axisymmetric model, even if $f_g = 0.15$ (see Fig. 3). The gas disks with $f_g \gtrsim 0.2$ fragments into several dense gas clumps due to the local gravitational instability. The Toomre's Q value is greater than one, which means gravitational stable for the axisymmetric linear perturbation, in $R < 1.5$ for $f_g = 0.10$, and $R < 0.7$ for $f_g = 0.15$, and $R < 0.05$ for $f_g = 0.30$. If the gas disk fragments into several clouds, it loses strong dissipative nature, and no rapid fueling occurs. As discussed in section 4.1, the dissipative nature of gaseous orbits plays a very important role in the fueling process.

### 3.2    Counter-rotating core formation

The accretion processes described in the previous subsection are basically common in models. However, we found in some models that rotation of the gaseous core finally formed in the center is in an *opposite* sense with respect to the outer gas and the galactic rotation as denoted in Table 1.

---

Fig. 6(a),(b)

---

Figure 6(a) and 6(b) are a position-velocity diagrams of model 37 and 40 in the inertial frame. Although the diagrams of two models are the almost same for the outer region, the nuclear cores show a different feature: the core in model 37 rotates in the same sense as the gas in the outer region, but in an opposite sense in model 40.

---

Fig. 7(a),(b); 8(a),(b)

---

In Fig. 7 and Fig. 8, velocity vectors of the gas with positive and negative angular momentum of the two models in the inertial frame are shown (i.e., prograde and retrograde rotation for the galactic rotation), respectively. We found that the dense core is formed earlier in model 37 ($T \sim 3.00$), which results in the positively rotation core, than in model 40 ($T \sim 3.43$). In other word, shocked gas in the more linear dense region in the massive models falls to the center in a shorter time scale. A negative angular momentum core is also formed, but the gas with positive one dominates in the core. On the other hand, the gas in a less massive one falls more slowly to the center. As seen in Fig. 7(b) and 8(b), a large fraction of gas in the linear shocked region get negative angler momentum by the torque from the weak bar potential at $T \sim 2.66 - 3.25$. Therefore, the dense shocked region become 'inverse S-shape' during the gases fall toward the center. That results in the retrograde core.

---

Fig. 9; 10

---

We plot total specific angular momentum in the inertial frame inside $R = 0.2$ for models with $\varepsilon_0 = 0.2$, as a function of the time (Fig. 9). There is a clear threshold of $f_g$ for the sense of the rotation: between 0.13 and 0.14 for the models with $\varepsilon_0 = 0.2$. This threshold tends to be smaller for weaker bar models ($\epsilon_0 = 0.1$ and 0.05). In Fig. 10, which is fraction of gas inside $R = 0.2$, we can see the counter-rotating core is formed in models with a smaller accretion rate (models 40, 41 and 42). We discuss physical reason of forming the retrograde core in section 4.2.

### 4    DISCUSSION

### 4.1    Why does the rapid fueling occur?

We have shown that a weakly distorted potential can induce the rapid gas supply into a few 10 pc region from a few kpc region in a massive gas disk ($\sim 5 - 20$ % of the dynamical mass inside the gas disk) in less than $\sim 10^7$. In this section we show that dissipational nature of the gaseous orbits and self-gravity of the gas are essential for the fueling process. Test particle orbits, that is, orbits without dissipation, in a barred potential have been analytically and numerically revealed. (Contopoulos & Papayannopoulos 1980). From these results, however, one can not explain the reason of rapid gas supply of our results. The characteristic behavior of the gas in a bar potential has been investigated by the epicycle theory as well as by hydrodynamical calculations in pioneering works (Sanders & Huntley 1976; Huntley, Sanders & Roberts 1978), and it has been found that gaseous orbits elongated at $45°$ in neighboring the Lindblad resonances. Recently, using a different analytical way, Wada (1994) has clearly and quantitatively shown gaseous behavior of a non-self-gravitating gas in a weak bar like potential by extending the epicycle approximation for the non-dissipational orbits.[*] On the rotating frame of the bar potential (angular velocity $= \Omega_b$), a linearized equation of motion of a particle with a friction, which oscillates with an amplitude $R_1$ around a guiding center at $R_0$, is

$$\ddot{R}_1 + 2\lambda \dot{R}_1 + \kappa_0^2 R_1 = f_0 \cos 2(\Omega_0 - \Omega_b) t, \quad (17)$$

where

$$f_0 \equiv - \left[ \frac{d\Phi_b}{dR} + \frac{2\Omega\Phi_b}{R(\Omega - \Omega_b)} \right]_{R_0}, \quad (18)$$

and $\lambda$ is the damping rate. From this equation of motion, we obtain a description of 'damped' closed orbits on the rotating frame of the bar,

$$R_1(\phi_0) = B \cos(2\phi_0 + \delta_0). \quad (19)$$

---

[*] Lindblad & Lindblad 1994 reached to almost the same conclusion.

Amplitude $B$ and the phase-shift, $\delta_0 \equiv \delta(R_0)$, are

$$B \equiv \frac{f_0}{\sqrt{\left\{\kappa_0^2 - 4(\Omega_0 - \Omega_b)^2\right\}^2 + 16\lambda^2(\Omega_0 - \Omega_b)^2}} \quad (20)$$

and

$$\delta_0 = \arctan\left[\frac{2F\Lambda}{F^2 - 1}\right], \quad (21)$$

where

$$F \equiv 2(\Omega_0 - \Omega_b)/\kappa_0, \ \Lambda \equiv \lambda/\kappa_0. \quad (22)$$

The important point is that the phase-shift $\delta_0$ is always negative if $F > 0$, and $\delta_0 = -\pi/2$ when $F^2 = 1$, that is, at the Lindblad resonances ($\Omega_0 = \Omega_b \pm \kappa_0/2$). The negative $\delta_0$ means that the damping oscillation is delayed from the periodic driven force. The phase delay of the damped oscillation appears as a leading shift of the oval orbit with respect to the bar potential if $B > 0$.

As can be seen in Figure 3 in Wada (1994), the amplitude $B$ is positive ($R < R_{\text{CR}}$) or negative ($R > R_{\text{CR}}$). Therefore the orientation of the major axis of the elliptical orbits to the bar major axis ($\Delta$) changes as follows:

$$0° < \Delta < 45° \quad \ldots\ldots \quad R_0 < R_{\text{ILR}_1}, \ R_{\text{ILR}_2} < R_0 < R_{\text{CR}},$$
$$\text{and } R_{\text{OLR}} < R_0,$$
$$45° < \Delta < 90° \quad \ldots\ldots \quad R_{\text{ILR}_1} < R_0 < R_{\text{ILR}_2}$$
$$\text{and } R_{\text{CR}} < R_0 < R_{\text{OLR}}.$$

Note that the oval gas orbits have always leading shift with respect to the bar potential inside the inner ILR. In fact, the shift is $30° \le \Delta \le 45°$ for a appropriate damping rate, $\lambda$, which is determined from a comparison between the theory and results of hydrodynamical simulations. Consequently the gas on the elongated orbits near to the ILRs loses its angular momentum (see Fig. 5 in Wada 1994). Both self-gravity and the angular momentum loss of the gas enhance the distortion of the oval orbits. Therefore the formation of the linear shocked region (Fig. 2) and the nuclear inflow of the gas inevitably occurs in a self-gravitating gas disk with the inner two ILRs of three ILRs, even if the distortion of the potential is very weak.

Let us consider here whether the second and the third ILR is necessary (see Fig. 1). In other words, what is the condition for the pattern speed to induce the effective fueling severe? Figure 6 in Wada (1994) shows that the phase shift of the orbits inside the corotation is not zero for the pattern speed that is two times larger than the critical one, for which there is one ILR. (Note that there are two or less ILRs in the potential in Wada (1994), since self-gravity of the gas disk was not consider.) This means that the oval orbits inside the corotation always incline in a leading sense not depending on the pattern speed.

Fig. 11

In Fig. 11, the phase shift of the oval orbit, $\delta_0$, derived from equations (4), (5), and (21), are plotted against the distance from the center for various pattern speed of the bar, taking into account the gravitational potential from the gas disk ($f_g = 0.15$). All cases show that $\delta_0$ is negative inside the initial radius of the gas disk ($R = 0.2$), and are less than about $-\pi/2$ at the vicinity of the center. (Remember that

$\delta_0$ is $-\pi/2$ at the ILRs.) Therefore the rapid fueling process found in models with $\Omega_b = 0.02$ could occur in a wide range of the pattern speed of the oval potential. Actually, model 49 ($\Omega_b = 0.03$), 52 and 57 ($\Omega_b = 0.04$), and 60 ($\Omega_b = 0.005$) show fueling process very similar to that in models with $\Omega_b = 0.02$ (Fig. 12).

Fig. 12

As expected the phase shift of the damped orbits (Fig. 11), orientation of the whole gas distribution in a fast rotating bar (model 57) is less than 45° with respect to the major axis of the bar, whereas it is larger than 45° in a slower bar (model 60).

As well as the pattern speed, mass of the gas disk is important constraint for causing the rapid fueling. Our results show that a very massive gas disk [model 23 and 21 ($f_g = 0.30$)] does not always induce the rapid fueling, since the gas disk fragments into clumps due to local gravitational instability. On the other hand, in the non self-gravitating gas disks ($f_g \ll 0.05$ for $\varepsilon_0 = 0.05$), collapse of the oval orbits at the center and the resultant fueling process does not occur, but the gas in an outer region accumulates to form a tilted oval ring. If the ring is gravitationally stable, it becomes aligned with the bar due to angular momentum exchange between the gas and the potential in $T \sim 30\text{-}40$ (see Fig. 3(b) in Paper I).

### 4.2 Why does the core counter-rotate?

As seen in Fig. 3, gas disk in a stronger bar more rapidly evolves and a larger amount of the gas can be fed into the center in more early phase. This fact implies that the torque exerted from the non-axisymmetric potential determines the evolution of the system. The torque, $\tau(R, \theta)$, caused by the bar potential (equation (2)) is

$$\tau(R, \theta) = \left(\frac{1}{R}\frac{d\Phi_1}{d\theta}\right) R$$
$$= -2\varepsilon a\alpha \frac{R^2}{(R^2 + a^2)^2}\sin 2\theta, \quad (23)$$

and has the extremum at $R = a$ and $\theta = \pm\pi/4, \pm 3\pi/4$.

Fig. 13; Fig. 14

If we assume the gas rotates anti-clockwise, the gas loses its angular momentum in the first and the third quadrant, as seen in Fig 13. Especially, the gas in the linear dense region leading by 45° to the bar potential most effectively looses its angular momentum. In Figure 14, trajectories of particles along the line inclined by 45° with respect to the non-axisymmetric potential, $\Phi(R, \theta) = \Phi_0 + \Phi_1$ are shown. The points, at which the initial angular momentum of the gas which rotates with the bar becomes zero, are shown for $\varepsilon_0 = 0.1$ and $0.2$. Fig. 15 shows that the gas in the inclined shocked region get negative angular momentum in a time scale $\sim 0.2$ and $0.4$ for $\varepsilon_0 = 0.1$ and $0.2$, respectively. This naive estimate implies that a core made from the gas, which is accumulated to the center along the 45° inclined line, can counter-rotate, if the gas gradually infalls to the center. More rapid infall is need for a stronger bar to keep the core positive rotation. Therefore dependence of the sense of rotation on the strength of the bar and mass of the gas disks can be reasonably understood.

### 4.3  Origin of the counter rotating core in galaxies

Kinematical decoupled stellar cores or a counter-rotating gaseous cores are often found in elliptical and S0 galaxies (Bender 1990; Rubin 1994). For example, IC 1459 has one of the strongest counter-rotating components of any observed elliptical galaxies (Forbes, Franx & Illingworth 1994). NGC 7252 has a counter-rotating molecular gas disk (Wang, Schweizer & Scoville 1992). The most widely accepted model for producing the counter-rotating cores is the retrograde merger of a compact elliptical galaxies (Kormendy 1984; Balcells & Quinn 1990). Our numerical results, however, give another possibility of forming the counter-rotating gaseous or stellar cores. We need an ovally distorted potential and a massive gaseous disk, which initially rotates with the same sense to that of the potential, in the inner region of galaxies (i.e., inside the turn over radius of the rotation curve). The rotating non-axisymmetric potential must be generated in most elliptical and S0 galaxies, which have a triaxial stellar distribution. Gaseous disks are not uncommon in the early-type galaxies, and must have existed in ellipticals if they were formed from mergers of gas rich spiral galaxies. Although star-formation should properly be taken into account in our numerical simulations, the observed stellar counter-rotating cores could be formed from the gaseous counter-rotating cores.

If it is correct, however, that the counter-rotating gaseous/stellar cores are originated from the resonant structure and the fueling process in a rotating bar, a questions arises: why gaseous and stellar counter-rotating cores have not been found in barred spiral galaxies and in spiral galaxies with a weakly distorted bulge? The probable reasons are (1) Rotating speed of most bar-like bulges in spiral galaxies is too high, or these galaxies were too rich for gas in the central region to cause the rapid fueling. The fraction of the gas mass ($f_g$ =0.03-0.12), and the pattern speed of the bar should be around the peak of $\Omega_0 - \kappa_0/2$. (2) The counter-rotating gaseous core found in our simulations is a transient phenomenon, which time-scale is $10^{7-8}$ yr. Because gas supply with positive angular momentum from the outer region could vanish the counter-rotating core. For example, gas accumulation to the ILR ring from a gas disk inside the corotation and its gravitational collapse (Fukunaga & Tosa 1991; Paper I; Friedli & Benz 1993) brings infall of a large amount of gas with positive angular momentum into the central region. This fueling process occurs in a later phase (several $10^8$ yr) than the rapid fueling process and the core formation. Spiral galaxies have a source of the positive angular momentum (i.e., outer disks), while ellipticals/S0 galaxies are poor in the outer gaseous disk. This difference might explain that E/S0 galaxies prefer the counter- rotating core.

### 4.4  Effects of the rapid fueling on the stellar bar

In our simulation, we neglected effects of gaseous evolution on the background bar-like potential. Since gas mass in the central region after the rapid fueling is comparable or sometimes greater than the dynamical mass that is derived from the background potential, we should take account of a response of the stellar bar to the self-gravitating gas. Numerical simulations with two components, i.e., stars and gases, have revealed that a gaseous component in a galactic disk affects on dynamical evolution in galaxies. Gaseous clumps caused by the local gravitational instability stir around the stellar disk and stabilize against the bar unstable mode (Shlosman & Noguchi 1993). Friedli & Benz(1993) have reported bar-dissolution due to bar-driven accretion of the gas into the central region. This phenomenon can be understood by an increase of chaotic orbits in the stellar bar (Hasan & Norman 1990). Our preliminary three dimensional SPH-$N$-body simulation, in which evolution of a gas disk in a rotating triaxial bulge-like stellar system was investigated, also revealed that the bar becomes to be weak after the gas accumulate to form a dense core in the center (Wada 1995, in preparation). This accretion process was very similar to that in the present paper, that is, a linear dense gas region leading the stellar bar was formed, and accreted to the center. Therefore we suspect that the back-reaction of the gaseous dynamics to the stellar system does not drastically change the rapid fueling in a fixed bar potential, because the destruction of the bar is occurred in typically several rotating period of the bar, which is longer enough than that of the rapid nuclear fueling.

Finally, we should note generality of our results, concerning especially to the initial conditions. Although the fueling process we have found is physically reasonable to occur, more initial conditions should be investigated in order to demonstrate generality of our conclusions. The effect of chosen initial conditions of the gas disk was not studied here and deserves a special attention. For example, the truncation of the disk between the 2nd and the 3rd ILR, which is because we adopted fine grids and therefore only a limited region was calculated, would be important in a longer time scale than that in the present simulations. We also note that existence of the effective fueling mechanism in massive gas disks suggests that the disk itself is unstable. One might doubt therefore that the initial massive gas disks we assumed are hardly formed, and the situations we have investigated here are astrophysically rare cases. However, the unstable disk appears only when the conditions for the rapid fueling are satisfied, for example, when the ILRs appear in a massive disk since the bar slows down due to the dynamical friction, or when the density of the gas disk exceeds the critical value by infalling the gaseous halo, merging of gas-rich satellites, or tidally induced accretion of gas from an outer disk by close-encounters of galaxies. Generality of these situations in galactic evolution needs further consideration.

### ACKNOWLEDGMENTS

We would like to thank B. G. Elmegreen for variable comments concerning the draft. We also acknowledge helpful discussions with I. Shlosman on several points of the paper. This work was supported in part by the Nobeyama Radio Observatory, Fellowships of the Japan Society for the Promotion of Science for Japanese Junior Scientists, Nukazawa

| Model | $\Omega_b$ (km s$^{-1}$ pc$^{-1}$) | $\varepsilon_0$ | $f_g$ | central core |
|-------|------|------|------|------|
| 26 | 0.02 | 0.0 | 0.15 | no |
| 56 | 0.02 | 0.05 | 0.05 | no |
| 55 | 0.02 | 0.05 | 0.08 | + |
| 47 | 0.02 | 0.05 | 0.10 | + |
| 45 | 0.02 | 0.05 | 0.12 | + |
| 22 | 0.02 | 0.05 | 0.15 | +(fragment) |
| 23 | 0.02 | 0.05 | 0.30 | no |
| 30 | 0.02 | 0.1 | 0.05 | − |
| 28 | 0.02 | 0.1 | 0.10 | − |
| 44 | 0.02 | 0.1 | 0.12 | + |
| 43 | 0.02 | 0.1 | 0.15 | + (fragment) |
| 21 | 0.02 | 0.1 | 0.30 | no |
| 54 | 0.02 | 0.2 | 0.03 | − |
| 41 | 0.02 | 0.2 | 0.10 | − |
| 50 | 0.02 | 0.2 | 0.11 | − |
| 40 | 0.02 | 0.2 | 0.12 | − |
| 42 | 0.02 | 0.2 | 0.13 | − |
| 36 | 0.02 | 0.2 | 0.14 | + |
| 37 | 0.02 | 0.2 | 0.15 | + |
| 38 | 0.02 | 0.2 | 0.16 | + |
| 39 | 0.02 | 0.2 | 0.17 | + |
| 60 | 0.005 | 0.2 | 0.15 | + |
| 49 | 0.03 | 0.2 | 0.15 | − |
| 52 | 0.04 | 0.2 | 0.15 | − |
| 57 | 0.04 | 0.2 | 0.17 | + |
| 53 | 0.06 | 0.2 | 0.15 | − |

**Table 1.** Model parameters and a sense of rotation of the nuclear core. In the fifth column, a plus sign means that a central core rotating in the same sense as the galactic rotation is formed, and a minus sign means retrograde cores. 'no' means the core is not formed.

## Figure Captions

### Fig. 1

$\Omega_0(R) - \kappa_0(R)/2$, from equations (4) and (5), for $f_g = 0.10$ (the thick line) and $f_g = 0.17$ (the thin line).

### Fig. 2

**(a)** Evolution of gas distribution of model 37. Major axis of the bar horizontally is fixed on each frame. Frame size is 5.1× 5.1. Time is shown at the right bottom corner.

**(b)**Same as Fig. 2(a), but velocity vectors of gas are shown. Unit vector ($V = 200$ km s$^{-1}$) is shown at the upper left corner. Frame size is 4.8×4.8.

**(c)** Same as Fig. 2(a) and (b), but close up of the central region. Frame size is 1.9×1.7. The core size at $T = 3.75$ is $0.09 \times 0.06$, and its mass is $10^9$.

### Fig. 3

Fraction of the gas mass to the total gas mass inside $R = 0.2$, as a function of time. Triangles, circles, and squares are models with $\varepsilon = 0.2, 0.1, 0.05$, respectively.

**Fig. 4**

Same as Fig. 2(b), but for model 22. Frame size is 5.4×5.4.

**Fig. 5**

(a) Isovelocity contour map is superposed on a density map of model 28 at $T = 5.87$. The gray scale represents surface density, $\log_{10}(\mu/\mu_0)$ , $\mu_0 \equiv 3 \times 10^5$. The position angle of the line of node is 135°. The inclination angle of the disk plane is assumed 30°. Major axis of the bar is horizontal.

(b) Same as Fig. 5(a), but for the position angle of the line of node is 45°.

**Fig. 6**

(a) Position-Velocity diagram of model 37 on an inertial frame. Angle between the line of sight and the bar major axis is 45° measured clockwise.

(b) Same as Fig. 6(a), but for model 40.

**Fig. 7**

Velocity vectors of gases with its specific angular momentum on an inertial frame a central region of model 37 (Fig. 7(a)), in which a positive rotating core is formed, and of model 40 (Fig. 7(b)), in which a negative one is formed. Time is shown at right bottom corner in each frame. The orthogonal solid line is position of the major axis of the oval potential. Frame size is 3.4×3.4.

**Fig. 8**

Same as Fig. 7(a) and 7(b), but for gases with negative angular momentum.

**Fig. 9**

Specific angular momentum (km s$^{-1}$ pc$^{-1}$) inside $R = 0.2$ of models with $\varepsilon_0 = 0.2$, as a function of time.

**Fig. 10**

Same as Fig. 9, but for the fraction of the gas mass to the total mass.

**Fig. 11**

Radial change of the phase shift of the damped orbit (equation (21)) taking into account of the gravitational potential of the gas disk ($f_g = 0.15$). Models with $\Omega_b = 0.01 - 0.025$ have three ILRs. The phase shift $\delta_0$ takes $-\pi/2$ at the ILRs.

**Fig. 12**

(a) Distribution of gas in the fast bar (model 57: $\Omega_b = 0.04$). Time is shown at the upper left corner. The major axis of the oval potential is horizontal.

(b) Same as Fig. 12(a), but for the slowest bar (model 60: $\Omega_b = 0.005$).

**Fig. 13**

Distribution of the torque $\tau(R, \theta)$ (equation (23)). Dashed contours represent the negative torque (we assume the rotation anti-clockwise). Contour are drawn from $-1$ to $+1$ every 1/16 (the unit is $-\varepsilon_0 \alpha/4a$).

**Fig. 14**

Trajectories of test particles along the line inclined by 45° in a potential (equation (1)). Distance from the center (y-axis) is plotted against time (x-axis). Points at which the initial angular momentum become zero are shown for $\varepsilon_0 = 0.2$ (open circles) and 0.1 (open squares).